\documentclass[12pt,letterpaper,onecolumn]{article}
\usepackage[latin1]{inputenc}
\usepackage{amsmath}
\usepackage{amsfonts}
\usepackage{amssymb}
\usepackage{braket}
\usepackage[left=1.00in, right=1.00in, top=1.00in, bottom=1.00in]{geometry}

\author{\\
\\
Horace P. Yuen\\
Department of Electrical Engineering and Computer Science\\
Department of Physics and Astronomy\\
Northwestern University, Evanston Il. 60208\\
yuen@eecs.northwestern.edu
}
\title{Fundamental Insecurity of Multi-Photon Sources Under Photon-Number Splitting Attacks in Quantum Key Distribution}

\begin{document}
\linespread{1}
\maketitle
\linespread{2}
\begin{abstract}
\textit{A  simple photon-number splitting attack is described which works on any lossy quantum key distribution system with a multi-photon source independently of the mean source photon number, and with no induced error rate. In particular, it cannot be detected by decoy states. The quantitative loss of security is similar when the user employs photon-number resolving detectors or threshold detectors. Numerical values indicate that existing implementations of concrete QKD systems are fundamentally insecure against this attack because a large portion of leaked sifted key bits is not accounted for. The possibility of other damaging photon-number splitting attacks is discussed. Some morals will be drawn.}
\end{abstract}
\newpage
\section{INTRODUCTION}

For BB84 type quantum key distribution (QKD) protocols, loss in the cryptosystem allows a powerful photon-number splitting (PNS) attack [1], whereby the attacker Eve splits off a photon from her multi-photon count upon her photon-number nondemolition measurement on the signal photon number. This has been dealt with extensively by the decoy state method [2-6], in which the user A sends out states of different average photon numbers $\langle n \rangle$ unknown to Eve during transmission at different bit intervals. The underlying idea is that Eve does not know $\langle n \rangle$ and cannot adapt her attack to states of different $\langle n \rangle$, typically one $\langle n \rangle$ is chosen to be the ``signal'' and the others are ``decoys''. In this way fundamental security is claimed to be preserved but with a much better ``secure key rate'' than otherwise [4--6]. In fact decoy states BB84 systems have been widely implemented. See for example [7--9].

It is well known and simple to see that Eve can split off any number of photons $m \leq n$ from the pulse once she identifies the number $n$ from her non-disturbing measurement, whether the system has polarization or other forms of state coding. (Note that in regard to the unconditional security claim associated with QKD, unless otherwise qualified we will consider any operation to be perfectly realizable by both the users A and B and the attacker Eve.) Indeed it is described in [10] how Eve may statistically reproduce, from an ``extended PNS attack'', a photon number distribution at the transmission channel output which is identical to what B would get for a Poisson source. This attack depends on knowing the mean $\langle n \rangle$ of the Poisson distribution, and thus is ineffective in the presence of decoy states.

In this paper, we will describe a specific photon-number splitting (SPNS) attack that would (statistically) reproduce the correct distribution at the channel output for any input source distribution, and dependent only on Eve's measured $n$. This attack cannot be detected by the users at all, including decoy state or any other source randomization strategy. Quantitatively this SPNS works much less effectively than the original PNS attack when the latter is not detected, but it is strong enough to place a severe limit on the signal multi-photon source. In particular the experimental systems of [7-9] are compromised by it even without any need for Eve to utilize a lossless link. Some implications of this SPNS attack on practical cryptosystem security and on fundamental security proofs will also be drawn.

\section{DIFFERENT PHOTON-NUMBER SPLITTING ATTACKS}

Consider a multi-photon source with probabilities $p_0$, $p_1$, and $p_m$ for emitting zero, one, and multiple photons during each pulsing or firing. In the presence of loss with transmittance $\eta$, the number of photons detected by the receiver user B occurs at a rate $\eta(p_1 + p_m)$, assuming each detection is a single count event which is the case for all existing single-photon detectors (often called ``threshold detectors''). Photon-number resolving detectors will be treated also in the following. Eve can split a photon off from of each of the $n \geq 2$ pulses, according to the usual  PNS attack, and transport the rest to the channel output by a lossless channel while deleting the $n = 1$ ones. Thus, when $p_1\eta \leq p_m$ security is totally compromised. This puts a stringent demand on $p_m$ being small. For a Poisson source with average photon number $\langle n \rangle$ obtained from a phase-randomized laser, this ratio $\frac{p_m}{p_1} \sim \langle n \rangle$ for small $\langle n \rangle$. This demand then makes the source emission rate $\sim \langle n \rangle$ similarly small, a huge efficiency problem. The same problem persists for ``single-photon sources'', which are characterized by the same two parameters $p, p_m$ and the actual physical firing rate, that these parameters are good.

Decoy states have been taken as a way to recover efficiency under this original PNS attack. Intuitively, if Eve carries out the above attack, she will stop more pulses with lower $\langle n \rangle$ which the users can check from the different $\langle n \rangle$ yields after B's detection. Eve's attack result is now characterized by the detected counts and quantum bit error rate (QBER) for each of the different $\langle n \rangle$ pulses. The typical recent conclusion [4--9] is that Poisson source of $\langle n \rangle \sim 0.5$ as the signal source among the decoy sources can be used with good security, similar to single-photon source.

That this cannot be true is simply seen as follows. Eve measures the $n$ of each pulse near the transmitter and passes all the single-count ones instead of blocking them! On multiple-count ones she splits off a photon and passes it along. For $\eta = 1$ and threshold counting, this would give no QBER but she learns a fraction
\begin{equation}
\frac{p_m}{(p_1 + p_m)}
\end{equation}
of B's detected photons. Numerically this is $\sim 23\%$ for a Poisson source with $\langle n \rangle \sim 0.5$. Decoy states would not detect this attack. This kind of leak is, however, \textit{not} accounted for in [4--9]. In the presence of loss, Eve could use lossless transport in principle to make up for the effect of her photon extraction at the channel output to preserve the original yield, but typically she wouldn't need to for such multi-photon loss effect. The large margin of uncertainty of the loss values for the transmission channel and B's optical components would make that unnecessary, especially with the fundamental inevitable statistical fluctuation associated with probabilities. Robustness of performance and false-alarm (aborting protocols when there is no need) probability are two crucial issues in concrete systems that have not been studied in QKD. In any event, such yield checking and also use of photon-number resolving detectors can be evaded as follows.

The following SPNS attack leads to no QBER and no difference in B's detected counts, assuming lossless channel replacement. Upon measuring $n$ on a pulse, Eve generates at the channel out $l$ photons according to the following binomial probability distribution or fraction of pulses:
\begin{equation}
B(n,l) = \binom{n}{l}\eta^{l}(1-\eta)^{n-l}
\end{equation}
Then B gets the following fraction of detected counts given $p_n$ is the source {\it n}-photon probability,
\begin{equation}
 p_1\eta + q^{B}_{m} = p_1\eta + \sum_{n=2}{p_n}\sum_{l=1}^{n}{B(n,l)}
\end{equation}

In (3), $q_{m}^{B}$ is B's probability of detecting various counts from the multi-photon part of the source from a photon-number resolving detector. Among B's detected photons, the fraction Eve gets is $q_{m}^{B}$ if B uses threshold detectors and is
\begin{equation}
q_{m}^{B} - \sum_{n=2}{p_{n}\eta^{n}}
\end{equation}
if B uses photon-number resolving detectors. This is because only in the latter case would Eve not have a photon from the multi-photon pulse with $n \geq 2 $. After public exchange Eve would know which photons in her possession are signal ones.

Decoy states is ineffective against this attack because the different $\langle n \rangle$ sources would be affected in the same way. (Statistical fluctuation from a finite number of pulses is a separate issue not analyzed.) Indeed, this attack falls outside the scope of the original PNS attack dealt with by decoy states. Note that the difference between (3) and (4) is typically negligibly small. Thus, Eve learns a fraction
\begin{equation}
\frac{q_{m}^{B}}{p_1\eta + q_{m}^{B}}
\end{equation}
of B's detected photons. In the limit of large loss $\eta \rightarrow 0$, (5) becomes
\begin{equation}
p_1 + p_m
\end{equation}
independent of $\eta$ for Poisson sources.

Generally, there are many possible photon-number splitting attacks. Eve can in principle get a list of all the $n$ values she measured from the pulses. With the specific given source randomization strategy and parameters, she could design her optimal attack and send back different photon numbers for each pulse accordingly. There may be many effective ways for her to deal with decoy states.

For example, from Bayes' rule Eve can tell, given states $\rho_i$ of different $\langle n \rangle_i$ employed with probability $\alpha_i$, the different probabilities $\lambda_i(n)$ that the measured $n$-pulse is from $\rho_i$. She can then delete it with an optimal resend strategy, in conjunction with other system imperfections in general. Indeed, statistically she can delete states of different $\langle n \rangle_i$ in the right proportion, for at least some parameter values, to escape decoy state detection. In large loss $\eta \ll 1$, security would be totally compromised again! This shows source randomization is essential. Detailed development will be given elsewhere. The essential point we would like to make here is that a general security proof \textit{cannot} be obtained without due consideration of \textit{all} of Eve's possible attacks.

\section{DECOY STATES AND UNCONDITIONAL SECURITY}

Since unconditional security is claimed on the basis of the decoy state method [4--9], a brief discussion is in order. In the original proposal [2], a specific (class of) PNS attacks is dealt with, but it is pointed out that more general attacks are possible and no unconditional security claim is made in [2,3]. The qualitative idea underlying decoy states is correct and useful, making it more difficult for Eve to delete single-photon counts. The important conclusion [2] that loss does not affect Eve's relative detected multi-photon rate is true also asymptotically for small $\eta$ in SPNS attacks, as (6) shows. It is an important open problem to ascertain whether a more general PNS attack, as described in section II, may re-introduce a serious unavoidable loss limit. It surely does if Eve knows the probability $\{\alpha_i\}$ for the states $\rho_i$, as indicated section II.

However, unconditional security from decoy states is claimed in [4--9]. In [4], the decoy state method is used with the number of measured counts and QBER, with asymptotic ``secure key rate'' established by the claim that, ``In principle, Alice and Bob can isolate the single-photon signals and apply privacy amplification to them only.'' This argument is a simplified version of the one in [11] on ``tagged'' photons and both are incorrect. There is no way the users can tell whether a detected single photon comes from a multi-photon pulse. Indeed, there would be no multi-photon problem if they could. Thus, the single-photon counts B uses would contain a fraction (5) from multi-photon pulses not accounted for in the security analysis.

Furthermore, there are several errors of omission in the ``security proof'' underlying such asymptotic key rates, including those of [5,6] where Eve's information gain from her knowledge of the error-correcting code and privacy amplification code are not accounted for. See also [12]--[13]. Here, in [4--6] the fraction of key bits from the SPNS attack is not properly accounted for in the security analysis. Whatever quantitative security claim hence asserted has no validity, especially given the large portion of sifted key bits that are so known to Eve, say in [7--9] discussed in the following.

Before we examine the numerical values, it may be mentioned that with the SPNS attack, any usual attack on the single-count states can be launched on top. In particular, some fraction of them can be subjected to intercept-resend attacks while keeping the QBER to any prescribed level. More importantly, Eve may design her joint attack with the knowledge that she will know exactly the bits from which she has splitted out a photon. It is not clear then how large a leak would prevent net key generation.

\section{IMPLICATIONS}

It is easy to see the consequence of this SPNS attack on concrete cryptosystems from (5) or just (1). For the NEC system [7], $p_0 \sim 0.6$, $p_1 \sim 0.3$, and $\eta \sim 4$ dB. This means 33\% of the sifted key bits are known to Eve. The same source applies to the Toshiba UK system [9] but with $\eta \sim 10$ dB. This means 38\% of the key bits are leaked. Note that (5) gives 40\% leak for $\eta \ll 1$, while (1) gives 23\% leak for the present source. The fact that this is not recognized and included in the security analysis suggests that the resulting security claim is totally unreliable. It is not clear a net key can still be generated after privacy amplification, especially under Eve's optimal attack.

One way to deal with SPNS attack is to use small $\langle n\rangle$ laser sources. However, as shown recently in [12], the tolerable QBER in QKD is actually very small, $\sim 1\%$ even just for collective attacks. Thus, a laser with $\langle n \rangle < 0.1$ needs to be used. Even then it is not yet known what Eve's optimum PNS attack is as discussed above. One may consider single-photon sources with small $p_m$, but one needs good $p_1$ and pulsing rate also. Another way is by the qb-KCQ system in [14], which also thwarts the known detector manipulation attacks as it turns out.

Loss is a major, perhaps \textit{the} major, problem in quantum information systems. Its possible effect on single-photon BB84 has been described in [15] and has not been seriously addressed in the literature. In particular, it is not practically possible to ascertain loss value in an optical system as complicated as a QKD system to any high degree of accuracy for purpose of security checks. The loss we have included in this paper is the transmission loss which Eve can in principle replace. There is always significant loss also in optical devices and components. Although we may assume Eve cannot manipulate such loss, she may take that into accounting in designing her attack. It is a daunting task to properly model a complete QKD system that depends so critically on these and other system parameter values. A more efficient and robust approach is needed.

\section{CONCLUSION}

A disconcertingly large number of papers, both on theory and on experiment, have made claims that the QKD system they discuss is unconditionally secure. In the case of theory, the claim sometimes is based on nothing more than a simple declaration, with perhaps a flimsy qualitative reason that is far from a proof. That is the case, for, example, on why a lossy channel merely changes the throughput but not the security of single-photon BB84, as discussed in [15]. Much effort has been spent on the decoy states approach with claimed unconditional security, but it simply does not give such guarantee as we show in this paper. The important point in this connection is not who have made errors, we all do. It is that, as discussed above, there simply has never been a proof offered. At best, only a specific kind of PNS attack is thwarted. Why can one then claim unconditional security? This is a major problem in almost every aspect of QKD security [12--16]. Since general security cannot be established by experiment, we have to deal with it much more seriously and critically.


\begin{thebibliography}{9}
\bibitem{ref1} N. Ginsin, G. Ribordy, W. Tittel, and H. Zbinden, Rev. Mod. Phys., 74, 145, 2002.
\bibitem{ref2} W. Y. Hwang, Phys. Rev. Lett. 91, 057901 (2003).
\bibitem{ref3} X. B. Wang, Phys. Rev. Lett. 94, 230503 (2005).
\bibitem{ref4} H. K. Lo, Z. Ma, and K. Chen, Phys Rev. Lett. 94, 230504 (2005).
\bibitem{ref5} M. Hayashi, Phys. Rev. A 76, 012329 (2007).
\bibitem{ref6} M. Hayashi, New J. Phys. 9, 284, 2007.
\bibitem{ref7} J. Hasegawa, M. Hayashi, T. Hiroshima, and A. Tomita, Asian Conference on Quantum Information Science, Kyoto, 2007; quant-ph 0705.3081.
\bibitem{ref8} Z. Yuan, A, Sharpe, and A. Shields, Appl. Phys. Lett. 90, 011118 (2007).
\bibitem{ref9} A. Dixon, Z. Yuan, J. Dynes, A. Sharpe and A. Shields, Appl. Phys. Lett. 96, 161102 (2010).
\bibitem{ref10} N. Lutkenhaus and M. Jahma, New J. Phys. 4, 44, 2002.
\bibitem{ref11} D. Gottesman, H.K. Lo, N. Lutkenhaus, and J. Preskill, Quantum Inf. Comput. 4, 325, 2004.
\bibitem{ref12} H. P. Yuen, arXiv: 1205.3820v.2, 2012.
\bibitem{ref13} H. P. Yuen, arXiv: 1205.0565v.2, 2012.
\bibitem{ref14} H. P. Yuen, IEEE J. Selected Topics in Quantum Electronics, 15, 1630 (2009).
\bibitem{ref15} H. P. Yuen, arXiv: 1109.1049, 2011.
\bibitem{ref16} H. P. Yuen, Phys. Rev. A 82, 062304 (2010).

\end{thebibliography}
\end{document}